\documentclass[3p]{elsarticle}

\usepackage{amssymb}
\usepackage{amsmath}
\usepackage{graphicx}
\usepackage{ucs}
\usepackage{url}
\usepackage[utf8x]{inputenc}
\usepackage{subfigure}
\usepackage{booktabs}

\journal{Physica A}

\begin{document}

\begin{frontmatter}



\title{Volume of the steady-state space of financial flows in a monetary 
stock-flow-consistent model }

\author[label1,label2]{Aurélien Hazan}
\address[label1]{Université Paris-Est}
\address[label2]{LISSI, UPEC, 94400 Vitry sur Seine, France}



\begin{abstract}
We show that a steady-state stock-flow consistent macro-economic model 
can be represented as a Constraint Satisfaction Problem (CSP).
The set of solutions is a polytope, which volume depends on the constraints
applied and reveals the potential fragility of the economic circuit,
with no need to study the dynamics. Several methods to compute the 
volume are compared, inspired by operations research methods and the
analysis of metabolic networks, both exact and approximate.
We also introduce a random transaction matrix, and study the particular
case of linear flows with respect to money stocks.
\end{abstract}


\end{frontmatter}


In this article we propose an approach to macro-economic modeling inspired by
stock-flow consistent (SFC) models \cite{godley_monetary_2007} and statistical physics,
solving a Constraint Satisfaction Problem (CSP) in a way similar to recent works in the field
of metabolic networks \cite{braunstein_estimating_2008}. 
The SFC framework provides accounting identities ensuring that "everything comes from
somewhere and everything goes some where"\cite[p.38]{godley_monetary_2007}, thanks to budget constraints and behavioral
constraints. 
The formalism of DSGE (Dynamic Stochastic General Equilibrium) is dominant 
today in macro-economics, partly because the corresponding models can be
written in the form of state-space models
and estimated in a well-studied statistical framework\footnote{~see \cite[§3.2]{fagiolo_macroeconomic_2012} for a discussion.}. 
Their usefulness has been widely debated among economists \cite{kirman_economic_2010,farmer_economy_2009} 
and physicists \cite{bouchaud_economics_2008} because of their inability to
predict crises.
Many of their hypotheses have been criticized, such as representative rational agents,
exogeneity of financial factors, clearing markets where offer always meet
demand, etc\ldots 
Moreover, DSGE models usually do not implement SFC accounting identities. 

Most SFC works take place at the macroeconomic aggregate level. Various assets 
(loans, equities, bonds, \ldots) and sectors (households, firms, banks, states,\ldots)
have been considered in the litterature \cite{caverzasi_post-keynesian_2015}. Models can be more or 
less detailed, depending on the focus of the study (for example, the production sector can be aggregated or multi-sectoral). 
The issue of the micro-foundations of SFC models has been tackled with the
combination of SFC and agent-based models (ABM).
ABM  \cite{tesfatsion_chapter_2006, dawid_agent-based_2008} can represent large populations of heterogeneous agents, to explore the influence 
of networks effects, coordination, bounded rationality and learning. However,
they do not usually implement stock-flow consistency.
Recent works combine SFC and ABM \cite{seppecher_flexibility_2012, kinsella_income_2011, bezemer_causes_2011}, providing
micro-foundations to SFC models, and imposing macro constraints to ABMs.
Nevertheless, the computational cost of ABM simulations is high, and theoretical understanding is limited so far.
Calibration and validation are known to be difficult problems.

We consider a simplified stock-flow consistent model developed by
macro-economists, where the state of the economy is the set of all stocks
and flows of money. It is shown 
that one can compute the set of admissible steady-state configurations 
of this simple model. In this steady-state solution space, all 
configurations are equally weighted, thus allowing unusual states 
of the financial flows to be encompassed. 
The marginal probabilities of individual configurations can be approximated
over the whole solution space. 

Our standpoint is to transpose ideas from the field of metabolic networks where
steady-state fluxes have been studied as CSP. These studies were in turn
inspired by Von Neumann's growth model of production economies \cite{von_neumann_model_1945}. 
The results obtained with metabolic networks were successfully compared to 
experimental data, as in the Red Blood Cell metabolism or the central metabolism
of E.coli \cite{wiback_monte_2004, braunstein_estimating_2008, orth_what_2010,guell_mapping_2015}.
Such system-scale studies reveal some interesting features of
metabolisms, for example the cooperation between pathways. 
It has been shown also that organisms such as E.coli do not necessarily optimize
their metabolic fluxes.
 
The steady-state equilibrium hypothesis is accepted in the field of metabolic
networks because of the separation of timescales between metabolic and genetic
regulations \cite{heinrich_metabolic_1978,palsson_systems_2006}. 
In economy, the existence of cycles and their corresponding time constants
has been the subject of many theories and debates. Recent empirical works 
are able to identify the timescale at which some specific phenomena operate
\cite{gallegati_wavelet_2014}.
In the case of the model examined in this article, we consider that at the
time scale that separates two balance sheets (one year), capital accumulation
and output growth are slow and will be considered constant (as noted
in \cite{piketty_capital_2013}, the global annual per capita growth rate 
of production is 0.8 \% on average on the 1700-2012 time interval).

The expected benefits of applying these methods in macro-economics include 
the analysis of fragilities, notably the sensitivity to arbitrary flow
constraints, such as shortages. Indeed, the volume of the solution space
evoked above is immediately impacted when constraints are added or removed,
and can reveal the flexibility or rigidity of financial flows subject to perturbations.
  
In section \ref{theory}, we detail the model of financial flows that will be used as a
benchmark, and present its background from a macro-economic modeling point 
of view. Comparisons are made with ABM and econophysics. We present
the different methods used to compute exactly and approximately the volume
of the steady-state solution space. Then in section \ref{results}, the experimental results 
are explained. Finally, sections \ref{discussion} and \ref{conclusion} 
are devoted to discussion and conclusion.

\section{Background and methods}
\label{theory}

\subsection{Steady-state solution space in a stock-flow-consistent model}
\label{sub.bmw model}

In SFC models agents are grouped by sectors (banks, firms,   
workers, state, central bank) that are linked by money transfers. For example:
\begin{itemize}
\item Banks lend money to firms, which pay interests to the former.
\item Banks pay interests on deposits made by workers. 
\item Firms pay wages to workers.
\item Workers buy consumption goods to firms.
\item Firms invest in capital goods bought from other firms.
\end{itemize}

Assets and liabilities at a given instant $t$ in time are summarized in a balance sheet, 
where positive and negative signs stand for uses and sources of money. 
The balance sheet in Tab. \ref{tab.balance} corresponds to the BMW model,
discussed in \cite[chap. 7]{godley_monetary_2007}, 
which will be used in this article. 
In the BMW model, bank issue loans to finance the investments of the productive
sector, while households are both consumers and workers. The state and central
bank are omitted. 
Production firms and the banks make no net profit. The net worth of these
sectors is zero. The net wealth equals the total tangible capital $K$. 

\begin{table}[htbp]
	\centering
	\begin{tabular}{lllll}
	\hline
	& Households & Production Firms & Banks & $\sum$ \\
	\hline
	Money deposits & +M & & -M &  0\\
	Loans & & -L & +L & 0 \\
	Fixed capital & & +K& & +K \\
	Balance (net worth) & -$V_h$ & 0  & 0 & -$V_h$ \\
	\hline
	$\sum$  & 0 & 0  & 0 & 0\\
	\hline
	\end{tabular}
\caption{\label{tab.balance} Balance sheet of the BMW model.
$M,L,K$ are the money deposits, loans, and tangible capital.
$V_h$ is the net worth of households.}
\end{table}

The transaction matrix sums up all the flows of funds between sectors 
within a time interval $[t,t+\Delta t]$. Positive and negative signs stand
for inflows and outflows of money. They are balanced using a double-entry 
book-keeping representation where rows sum to zero since each transaction
has a counterparty, and columns sum to zero because of the sector's budget constraints.

Tab. \ref{bmw.transactions} shows the transaction matrix
corresponding to the BMW model with one agent per sector.
After \cite[chap. 7]{godley_monetary_2007},
we make the hypothesis that demand terms, with the subscript $d$, equal supply terms,
with the subscript $s$. Notations are summarized in Tab. \ref{bmw short labels}.

\begin{table}[htbp]
\centering
	\begin{tabular}{lllllll}
	\hline
	&Households&\multicolumn{2}{c}{Production Firms}& \multicolumn{2}{c}{Banks} &\\
	\cmidrule(lr{.75em}){3-4}  \cmidrule(lr{.75em}){5-6}
	 &  & Current& Capital & Current & Capital & $\sum$ \\
	\hline
	Consumption & $-C_d$ & $C_s$ & &  &  & $0$ \\
	Investment & & $I_s$ & $-I_d$ & & & $0$ \\
	Wages & $WB_s$ & $-WB_d$ & & & & $0$ \\
	Depreciation  & & $-AF_d$ & $AF_s$ & & & $0$ \\
	Interest on loans   & & $-IL_d$ & & $IL_s$ & & $0$ \\
	Interest on deposits  & $ID_s$ & & & $-ID_d$ &  & $0$ \\
	\hline
	Change in loans   & & & $\Delta L$  & & -$\Delta L$ & $0$ \\
	Change in deposits   & -$\Delta M$ & & & & $\Delta M$ & $0$ \\
	\hline
	$\sum$  & $0$ & $0$ & $0$ & $0$  & $0$ & $0$\\
	\hline
	\end{tabular}
\caption{\label{bmw.transactions}Transaction matrix of the BMW model. In
the stationary case, $\Delta L=\Delta M=0$.}
\end{table}

In the time-dependent case, the balance sheet at time $t+\Delta t$ is obtained 
adding stocks $M,L$ at time $t$ to changes in deposits and loans
$\Delta M$ and $\Delta L$ that occurred during the time interval $[t,t+\Delta t]$.

Since in this article the stationary steady-state case is considered,
the balance sheet and the transaction matrix are constants, and
the change terms $\Delta M, ~\Delta L$ are set to zero.

Furthermore in the BMW model a set of behavioral equations expresses several
flows as linear functions of the other variables and of fixed parameters.
The demand for consumption in equation Eq. (\ref{eq.cons})
depends on the disposable income and on the money deposit of the household:
\begin{eqnarray}
C_d &=&  \alpha_0 + \alpha_1 YD + \alpha_2 M \nonumber \\
    &=& \alpha_0 + \alpha_1 (WB_s + r M) + \alpha_2 M \label{eq.cons} 
\end{eqnarray}
where $\alpha_0,~\alpha_1,~\alpha_2$ are consumption parameters, and $YD$ is the
disposable income. 
The depreciation of tangible capital is proportional to its stock:
\begin{eqnarray}
AF &=& \delta K 
\end{eqnarray}
where $\delta$ is the rate of depreciation.  
Interests are proportional to stocks:
\begin{eqnarray}
IL_s &=& r L \label{eq.interest.loan} \\
ID_s &=& r M \label{eq.interest.deposit}
\end{eqnarray}
where the interest rate $r$ is the same for deposits and loans, for simplicity.

The differents constraints (row sums, columns sums, behavioral equations,
demand equals supply) can be written in matrix form: 
\begin{equation}
S = \{ x ~\text{s.t.} ~\xi x =b,\forall i ~x_i \in [0,x_i^{\max}] \} \\
\label{eq.matrix}
\end{equation}
where $x_{i}^{\max}$ sets the maximum flow for each component, and: 
\begin{equation}
x= \begin{bmatrix}
C_s & C_d & I_s & I_d & WB_s & WB_d & AF_s & AF_d & IL_s & IL_d & ID_s & ID_d  \\
\end{bmatrix}^T
\label{eq.def.x.1D}
\end{equation}
and:
\begin{equation}
\xi = \begin{bmatrix}
1 &-1 & 0 & 0 & 0 & 0 & 0 & 0 & 0 & 0 & 0 & 0\\
0 & 0 & 1 &-1 & 0 & 0 & 0 & 0 & 0 & 0 & 0 & 0\\
0 & 0 & 0 & 0 & 1 & -1& 0 & 0 & 0 & 0 & 0 & 0\\
0 & 0 & 0 & 0 & 0 & 0 & 1 & -1& 0 & 0 & 0 & 0\\
0 & 0 & 0 & 0 & 0 & 0 & 0 & 0 & 1 & -1& 0 & 0\\
0 & 0 & 0 & 0 & 0 & 0 & 0 & 0 & 0 & 0 & 1 &-1\\
0 &-1 & 0 & 0 & 1 & 0 & 0 & 0 & 0 & 0 & 1 & 0\\
1 & 0 & 1 & 0 & 0 & -1& 0 & -1& 0 &-1 & 0 & 0\\
0 & 0 & 0 & -1& 0 & 0 & 1 & 0 & 0 & 0 & 0 & 0\\
0 & 0 & 0 & 0 & 0 & 0 & 0 & 0 & 1 & 0 & 0 &-1\\
0 & -1 & 0 & 0 & \alpha_1 & 0 & 0 & 0 & 0 & 0 & 0 & 0\\
0 & 0 & 0 & 0 & 0 & 0 & 0 &-1 & 0 & 0 & 0 & 0\\
0 & 0 & 0 & 0 & 0 & 0 & 0 & 0 & 0 & -1& 0 & 0\\
0 & 0 & 0 & 0 & 0 & 0 & 0 & 0 & 0 & 0 & 1 &0\\
\end{bmatrix}
\end{equation}
Then:
\begin{equation}
b = \begin{bmatrix}
0 & 0&0&0&0&0&0&0&0&0& -\alpha_0-(\alpha_1 r+\alpha_2 ) M & -\delta K & -L & rM\\
\end{bmatrix}^T
\end{equation}

$\xi$ is an $m \times n$ matrix with $m=14$ and $n=12$, $m$ being the 
number of equations, and $n$ the number of unknown flows.
The first six rows of $\xi$ represent the row sums constraints of the
transaction matrix. The following four rows represent the budget constraints.
The last four rows are equivalent to the behavioral equations. 
Closed form solutions of $S$ are studied in \cite[chap. 7]{godley_monetary_2007}. 

Constraints can be added or modified: for example, if a firm goes bankrupt,
the loan will not be completely repaid, thus equation $IL_s=rL$ may be replaced by 
$IL_s \leq rL$. Agents may spend less in consumption than what is
prescribed by the behavioral equation, which can be written 
$C_d \leq \alpha_0 + \alpha_1 YD + \alpha_2 M$.

In section \ref{sub.manyagents}, the scope of the model is extended to the case
of multiple agents per sector. In that case, the linear system will be under-determined.

\begin{table}[htbp]
\centering
\begin{tabular}{ll}
\hline
 Variable & Label\\
\hline
 money deposit & $M$ \\
 capital & $K$ \\
 loans to firms  & $L$   \\
 investment & $I$  \\
 interest on loans & $IL$  \\
 wage bill & $WB$   \\
 depreciation allowance & $AF$  \\
 interest on workers deposits & $ID$  \\ 
 consumption of workers & $C$  \\
\end{tabular}
\caption{Labels associated with the different monetary variables, after 
\cite{godley_monetary_2007}. The subscripts $d$ and $s$ stand for demand and supply. }
\label{bmw short labels}
\end{table}

\subsection{Many agents per sector}
\label{sub.manyagents}

To extend the model of section \ref{sub.bmw model} to the case of many agents per sector,
the scalars in matrix $\xi$ will be replaced by block matrices.
The number of banks, firms and workers are noted $nb,nf,nw$.
Each block will be designed to account not only for the flows and stocks 
but also for the connectivity between agents.

For example, the first row of matrix $\xi$, $[1,-1,0, \ldots,0]$, may be written
with blocks instead of scalars, in order to encode the relationship 
between demand and supply of consumption goods. 
The positive term, that corresponds to the supply side of consumption, can be replaced 
by the identity matrix  $I_{nf}$. The negative term, that corresponds to the demand side,
can be replaced for example by:
\begin{equation}
\mathbf{A}_{nf,nh}=\begin{bmatrix}
0      &  1     & \ldots & 0      \\
0      &  0     &        & 0      \\
\vdots & \vdots &        & \vdots \\
0      &  0     &        & 0      \\ 
1      &  0     & \ldots & 1
\end{bmatrix},
\end{equation}
$\mathbf{A}_{nf,nh}$ is an $n_f \times n_h$ matrix, such that $\mathbf{A}_{nf,nh}[i,j]=1$
if the firm $i$ is selling consumption goods to the household $j$. In this example,
the first and the last households are clients of the same firm.

We suppose that one household can buy consumption goods from one firm only, chosen 
uniformly at random among the $n_f$ firms, and that firms can sell goods
to many households, without restriction. Thus the block matrices
$\mathbf{A}_{nf,nh}$ will be randomly sampled from the set $\{0,1\}^{n_f \times n_h}$ such
that columns sum to $1$.

Similarly, the following random matrices are introduced:
\begin{itemize}
\item $\mathbf{B}_{nf,nf}$ encodes investements: firms buy capital goods 
from one firm only, $\mathbf{B}_{nf,nf}[i,j]=1$ if the firm $j$ is selling
capital goods to the firm $i$. $\mathbf{B}_{nf,nf}$ is randomly sampled
from the set $\{0,1\}^{n_f \times n_f}$ with column sums and row sums equal to $1$.
\item $\mathbf{C}_{nf,nh}$ encodes wages: one firm pays wages to many households,
households get a wage from one firm only. $\mathbf{C}_{nf,nh}[i,j]=1$ if firm $i$ 
pays a wage to household $j$.  $\mathbf{C}_{nf,nh}$ is randomly sampled
from the set $\{0,1\}^{n_f \times n_h}$ with column sums equal to $1$.
\item $\mathbf{D}_{nb,nf}$ encodes interests on loans: one bank grants 
loans to many firms which pay interests in return, firms get loans from one bank only,
and pay interests to this bank only. 
$\mathbf{D}_{nb,nf}[i,j]=1$ if bank $i$ is being paid interests by firm $j$.
$\mathbf{D}_{nb,nf}$ is randomly sampled from the set $\{0,1\}^{n_b \times n_f}$ 
with column sums equal to $1$.
\item $\mathbf{E}_{nb,nh}$ encodes interests on deposits: each household keeps their deposit
on one bank account. Banks pay interests in return, and have many accounts opened
for their clients. $\mathbf{E}_{nb,nh}[i,j]=1$ if bank $i$ pays interests to 
household $j$. $\mathbf{E}_{nb,nh}$ is randomly sampled from the set 
$\{0,1\}^{n_b \times n_h}$ with column sums equal to $1$.
\end{itemize}

In Tab. \ref{bmw balance many} we give an example of balance sheet extended to 
the case of many agents per sector, with random connectivity as explained above.
The associated transaction matrix is written in Tab. \ref{bmw transac many}.
In the latter, an example of random choice by households of
the firm they buy goods from can be observed. The connectivity of flows of interests 
and changes in loans and deposits respects the one randomly defined in Tab. \ref{bmw balance many}.

\begin{table}[htbp]
	\centering
	\begin{tabular}{lllllllll} 
	\hline
	& \multicolumn{3}{c}{Households}  & \multicolumn{2}{c}{Firms} & \multicolumn{2}{c}{Banks} & $\sum$ \\
	\cmidrule(lr{.75em}){2-4}  \cmidrule(lr{.75em}){5-6} \cmidrule(lr{.75em}){7-8}
	 & 1 & 2 & 3 & 1 & 2 & 1 & 2 &\\
	\hline
	Money deposits & $M_1$  &  &  &  &  &  & -$M_1$ & 0\\
		           &        & $M_2$  &  &  &  & -$M_2$ &  & 0\\	
		           &        &  & $M_3$  &  &  &  & -$M_3$ &  0 \\	
	\hline
	Loans  &   &  &  & -$L_1$ &  & $L_1$ &  & 0\\		           
	       &   &  &  &  & -$L_2$ &  & $L_2$ & 0\\		           
	\hline
	Fixed capital  &   &  &  & $K_1$ &  & &  & $K_1$\\		           
	               &   &  &  &  & $K_2$  & &  & $K_2$\\		           
	\hline               
	Balance (net worth)  & -$V_{h1}$  & -$V_{h2}$ & -$V_{h3}$ &  &  & &  & -$\sum_i V_{hi}$ \\		                          		
	\hline
	$\sum$               & 0  &0  &0  & 0 &0  &0 & 0 & 0\\		        
	\hline
	\end{tabular}
\caption{\label{bmw balance many} Example of balance sheet of the BMW 
model with many agents $nw=3$, $nf=2$, $nb=2$.
Households and firms randomly choose their bank.
$M_i,L_j,K_k$ are the individual money deposits, loans, and tangible capital.
$V_{hi}$ is the net worth of individual households.}
\end{table}

\begin{table}[htbp]
\centering
	\begin{tabular}{lllllllllllll} 
	\hline		
 	& \multicolumn{3}{c}{Households}  & \multicolumn{4}{c}{Production Firms} & \multicolumn{4}{c}{Banks} & $\sum$ \\
 	\cmidrule(lr{.75em}){2-4}  \cmidrule(lr{.75em}){5-8} \cmidrule(lr{.75em}){9-12}
	&   &   &   & \multicolumn{2}{c}{Current} & \multicolumn{2}{c}{Capital}  & \multicolumn{2}{c}{Current} & \multicolumn{2}{c}{Capital}  & \\
	\cmidrule(lr{.75em}){5-6}  \cmidrule(lr{.75em}){7-8} \cmidrule(lr{.75em}){9-10} \cmidrule(lr{.75em}){11-12}
	& 1 & 2 & 3 & 1 & 2 & 1 & 2 & 1 & 2 & 1 & 2 & \\
	\hline
	Consumption & -$C_{d1}$ &  &  & $C_{d1}$ &  &  &  &  &  &  &  & 0\\
	            &  & -$C_{d2}$ &  & $C_{d2}$ &  &  &  &  &  &  &  & 0\\
	            &  &  & -$C_{d3}$ &  & $C_{d3}$  &  &  &  &  &  &  & 0\\
	\hline            
	Investment &  &  &  &  & $I_{s1}$ & -$I_{s1}$ &  &  &  &  &  & 0\\            
	           &  &  &  & $I_{s2}$ &  &  & -$I_{s2}$ &  &  &  &  &0 \\            
	\hline            	
	Wage & $WB_{s1}$ &  &  &  & -$WB_{s1}$ &  &  &  &  &  &  & 0\\
	     &  & $WB_{s2}$ &  & -$WB_{s2}$ &  &  &  &  &  &  &  & 0\\
	     &  &  & $WB_{s3}$ & -$WB_{s3}$ &  &  &  &  &  &  &  & 0\\
	\hline            	
	Depreciation &  &  &  & -$AF_1$ &  & $AF_1$ &  &  &  &  &  & 0\\            
	             &  &  &  & & -$AF_2$  &  & $AF_2$ &  &  &  &  & 0\\            
	\hline            	
	Interest on loans &  &  &  & -$IL_1$ &  &  &  & $IL_1$ &  &  &  & 0\\            
					  &  &  &  &  & -$IL_2$ &  &  &  & $IL_2$ &  &  & 0\\            
	\hline            	
	Interest on deposits & $ID_1$ &  &  &  &  &  &  & &  -$ID_1$ &  &  & 0\\            
					     &  & $ID_2$ &  &  &  &  &  & -$ID_2$ &  &  &  & 0\\            
						 &  &  & $ID_3$  &  &  &  &  &  & -$ID_3$ &  &  & 0\\            	
	\hline
	Change in loans &  &  &  &  &  & $\Delta L_1$ &  & &   & -$\Delta L_1$  &  & 0\\            
	                &  &  &  &  &  &  & $\Delta L_2$ & &   &   & -$\Delta L_2$ & 0\\            
	\hline
	Change in deposits & -$\Delta M_1$ &  &  &  &  &  &  & &   &   & $\Delta M_1$ & 0\\            
	                   &  & -$\Delta M_2$ &  &  &  &  &  & &   &  $\Delta M_2$ &  & 0\\            
	                   &  &  & -$\Delta M_3$ &  &  &  &  & &   &   & $\Delta M_3$ & 0\\            	
	\hline
	$\sum$ & 0 & 0 & 0 & 0 & 0 & 0 & 0 & 0 & 0   & 0  & 0 & 0\\            
	\hline
	\end{tabular}
\caption{\label{bmw transac many}Example of transaction matrix of the BMW model 
with many agents $nw=3$, $nf=2$, $nb=2$. 
Households randomly choose firms to buy consumption goods. 
Firms randomly choose other firms to buy capital goods, and the workers they hire.
Interests and changes in loans and deposits are set according to the balance sheet. 
In the stationay case, $\Delta L=\Delta M=0$.}
\end{table}

The following constraint satisfaction problem in matrix form sums up 
the set of constraints resulting from the balance sheet in Tab. \ref{bmw balance many},
the transaction matrix in Tab. \ref{bmw transac many},
and the behavioral equations in eq.(\ref{eq.cons}-\ref{eq.interest.deposit}):

\begin{equation}
S=\{ x ~s.t. ~\xi x=b , \forall i ~x_i\in [0,x_{i}^{\max}]  \}
\label{eq.matrix.many}
\end{equation}
where:

\begin{equation}
\xi = \begin{tiny}\begin{bmatrix}
I_{nf} &-\mathbf{A} & 0 & 0 & 0 & 0 & 0 & 0 & 0 & 0 & 0 & 0 & 0 &0 &0\\
0 & 0  & -\mathbf{B} & -I_{nf} & 0 & 0 & 0 & 0 & 0 & 0 & 0 & 0 & 0 &0 &0  \\
0 & 0 & 0 & 0  & \mathbf{C} & -I_{nf} & 0 & 0 & 0 & 0 & 0 & 0 & 0 &0&0 \\
0 & 0 & 0 & 0 & 0 & 0  & I_{nf} & -I_{nf} & 0 & 0 & 0 & 0 & 0 & 0 & 0 \\
0 & 0 & 0 & 0 & 0 & 0 & 0 & 0  & I_{nb} & -D & 0 & 0 & 0 & 0 & 0  \\
0 & 0 &0 & 0 & 0 & 0 & 0 & 0 & 0 & 0  & I_{nb} & -E & 0 & 0 & 0 \\
0 &-I_{nh} & 0 & 0 & I_{nh} & 0 & 0 & 0 & 0 & 0 & I_{nh} & 0 & 0 & 0 &0\\
I_{nf} & 0 & I_{nf} & 0 & 0 & -I_{nf}& 0 & -I_{nf}& 0 &-I_{nf} & 0 & 0 &0 &0 &0\\
0 & 0 & 0 & -I_{nf}& 0 & 0 & I_{nf} & 0 & 0 & 0 & 0 & 0 & 0 & 0 & 0\\
0 & 0 & 0 & 0 & 0 & 0 & 0 & 0 & I_{nb} & 0 & 0 &-I_{nb} & 0 & 0 & 0\\
0 & -I_{nh} & 0 & 0 & \alpha_1 I_{nh} & 0 & 0 & 0 & 0 & 0 & 0 & 0 & -(\alpha_1 r +\alpha_2) I_{nh} &0&0\\
0 & 0 & 0 & 0 & 0 & 0 & 0 &I_{nf} & 0 & 0 & 0 & 0 &0&0&-\delta I_{nf}\\
0 & 0 & 0 & 0 & 0 & 0 & 0 & 0 & 0 & I_{nf} & 0 & 0 &0 &-r I_{nf} &0\\
0 & 0 & 0 & 0 & 0 & 0 & 0 & 0 & 0 &  0 & I_{nh} &  0 & -r I_{nh} &0 &0 \\
0 & 0 & 0 & 0 & 0 & 0 & 0 & 0 & 0 &  0 & 0 &  0 & 1\hdots 1 &0 &0 \\
0 & 0 & 0 & 0 & 0 & 0 & 0 & 0 & 0 &  0 & 0 &  0 & 0 &0 & 1\hdots 1 \\
0 & 0 & 0 & 0 & 0 & 0 & 0 & 0 & 0 &  0 & 0 &  0 & 0 & I_{nf} & -I_{nf} \\
0 & 0 & 0 & 0 & 0 & 0 & 0 & 0 & 0 &  0 & 0 &  0 & -E & D & 0 \\
\end{bmatrix}
\end{tiny}
\end{equation}

Rows 1 to 6 of $\xi$ express the row sums constraints of the transaction matrix.
Rows 7 to 10 represent the column sums. Rows 11 to 14 are equivalent to the behavioral
equations. The last rows correspond to the balance sheet.

Compared to eq. (\ref{eq.def.x.1D}), $x$ now includes the individual stocks $M_i,L_j,K_k$:
\begin{equation}
x= \begin{bmatrix}
C_s^T & C_d^T & I_s^T & I_d^T & WB_s^T & WB_d^T & AF_s^T & AF_d^T & IL_s^T & IL_d^T & ID_s^T & ID_d^T & M^T &L^T &K^T \\
\end{bmatrix}^T
\end{equation}
where $x$ is written as a concatenation of vectors. As an example, 
$M^T=[M_1,\ldots, M_{nh}]^T$.  We also write:
\begin{equation}
b= \begin{bmatrix}
0_{nf} & 0_{nf} & 0_{nf} & 0_{nf} & 0_{nb} & 0_{nb} &0_{nh}&0_{nf}&0_{nf}&0_{nb} & \alpha_0.1_{nh}  & 0_{nf}&0_{nf}&0_{nh}& M_{\textrm{tot}}& M_{\textrm{tot}}& 0_{nf} &  0_{nb}\\
\end{bmatrix}^T
\end{equation}
where $0_{n}$ and $1_{n}$ are vectors, and $M_{\textrm{tot}}$ is the total quantity
of deposits.

\begin{figure}[htbp]
\centering
\includegraphics[width=6cm]{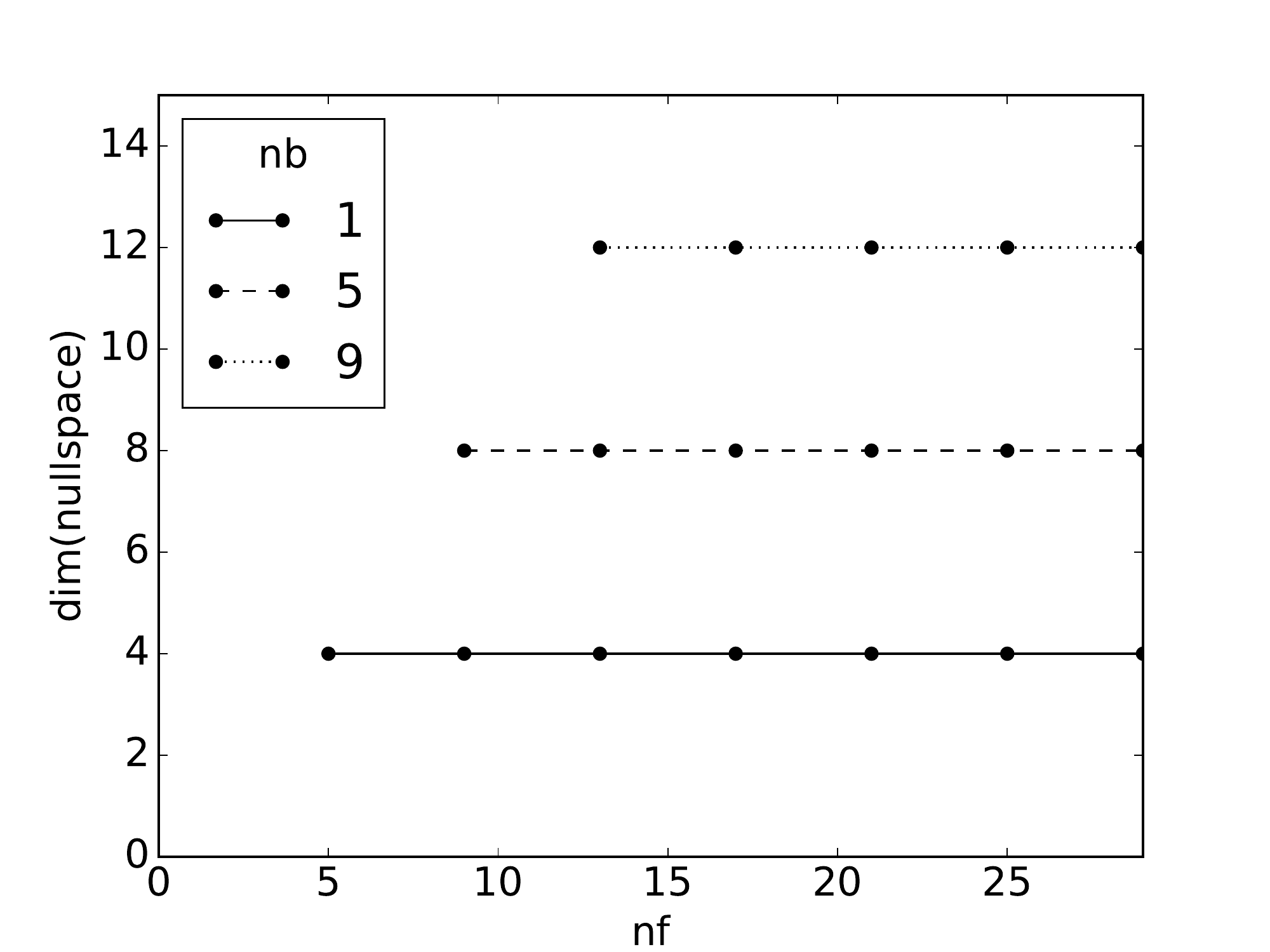}
\caption{Dimension of the nullspace of $\xi$ determined numerically
with $r=0.03, ~\alpha_1=0.75, ~\alpha_2=0.1, ~\delta=0.1$, and $nw=100$. 
This dimension is computed only for values of $nf$ identified by the marker $\bullet$. $nb,nf,nw$ are such that $nb<nf<nw$
}
\label{fig.nullspace}
\end{figure}

The dimension of the nullspace of $\xi$ is then determined numerically 
for a specific set of parameters consistent with \cite{godley_monetary_2007},
and such that $nb<nf<nw=100$. 
We find that in the cases examined in Fig. \ref{fig.nullspace},
the matrix $\xi$ doesn't have full rank, and that consequently
the system $\xi x=b$ is under-determined.

The upper bounds $x_{i}^{\max}$ are set independently for each variable, and are
logic consequences of the positivity of the flows, and of the conservation
of the total stock of money :
\begin{itemize}
\item $\forall i \in [1,nh], M_i \in [0,M_{\textrm{tot}}]$. 
\item $\forall i \in [1,nf], L_i \in [0,M_{\textrm{tot}}]$.
\item $\forall i \in [1,nf], K_i \in [0,M_{\textrm{tot}}]$.
\item $\forall i \in [1,nf], AF_i \in [0,\delta M_{\textrm{tot}}]$.
\item $\forall i \in [1,nf], I_i \in [0,\delta M_{\textrm{tot}}]$.
\item $\forall i \in [1,nh], ID_i \in [0,r M_{\textrm{tot}}]$.
\item $\forall i \in [1,nf], IL_i \in [0,r M_{\textrm{tot}}]$.
\end{itemize}

\begin{figure}[htbp]
\centering
\subfigure[~]{
	\includegraphics[width=6cm]{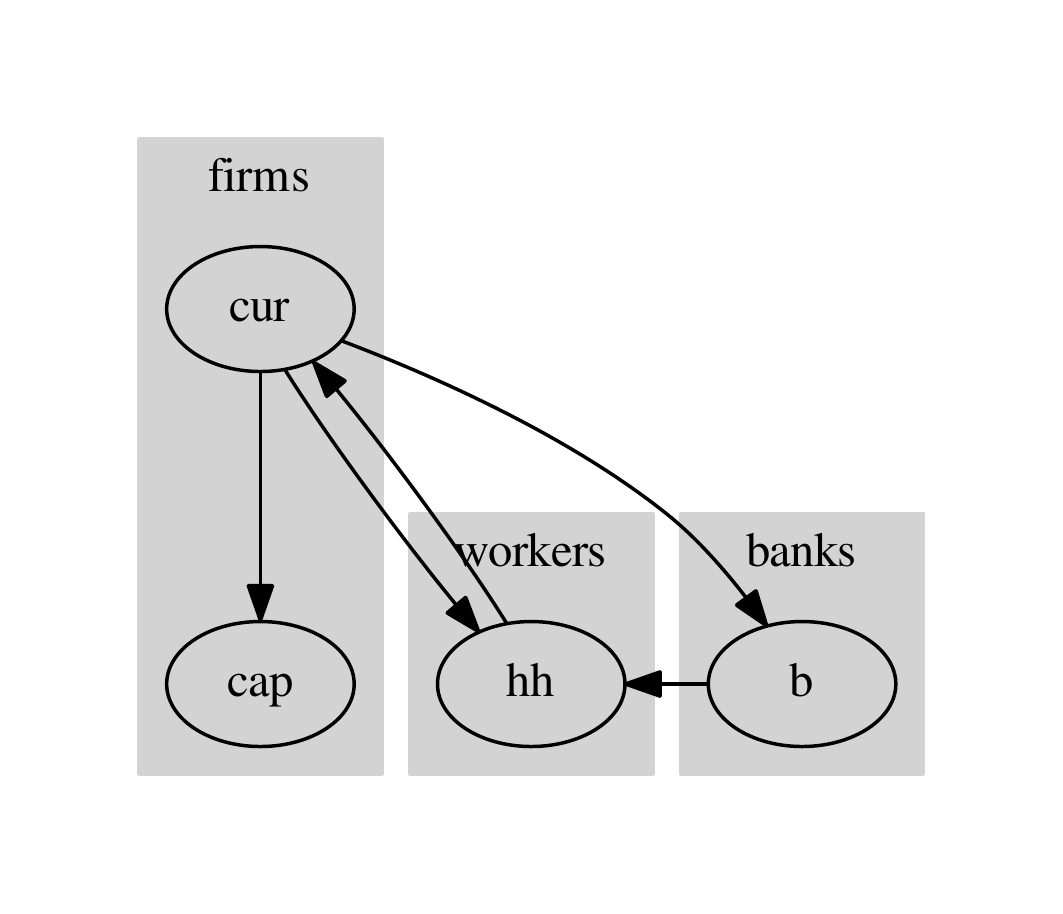}}
\subfigure[~]{
	\includegraphics[width=8cm]{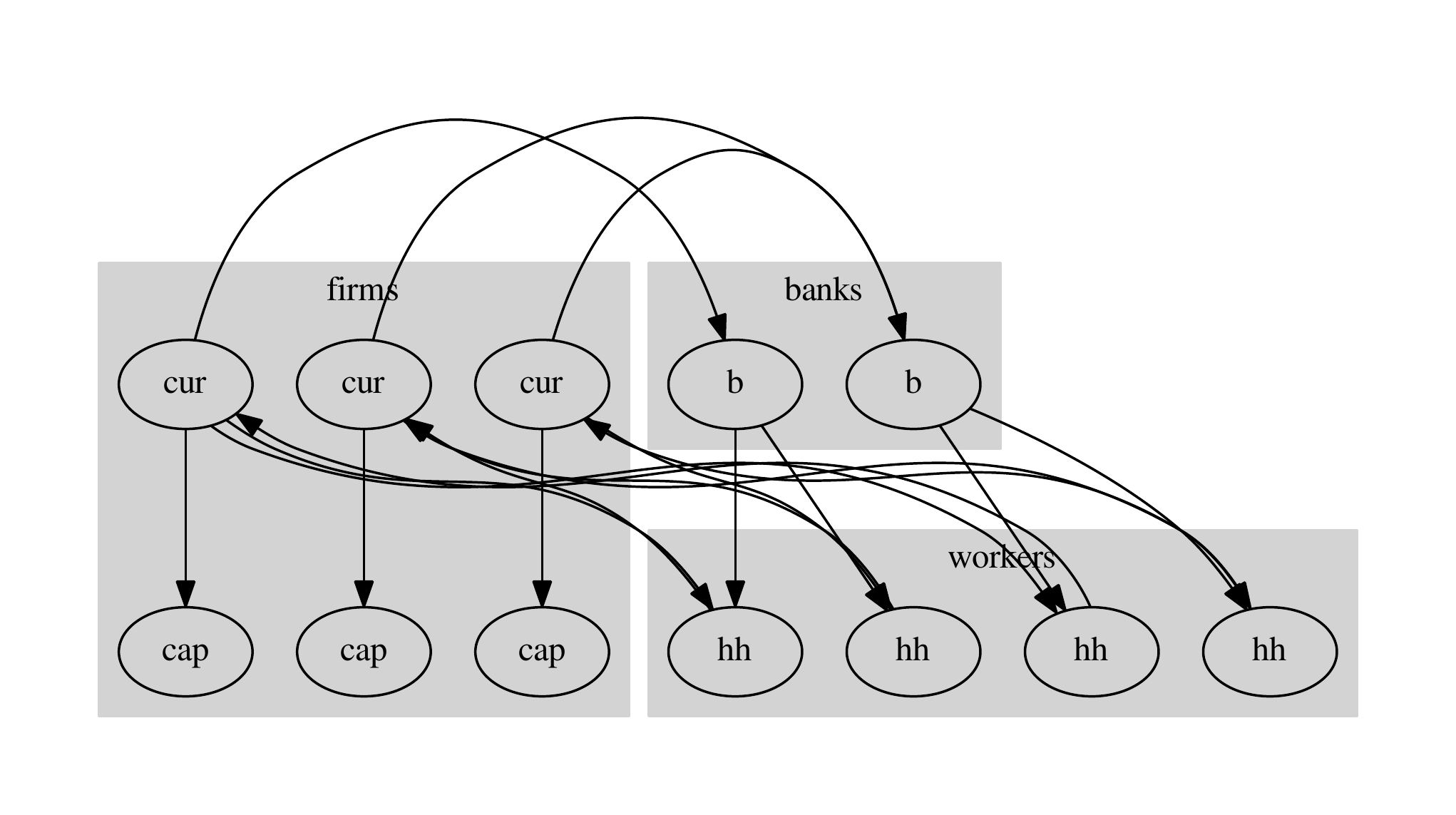}}
\caption{Graph of a subset of monetary transactions.
\emph{cur,cap,w,b} stand for firm's current and capital accounts, worker's deposit
 accounts, bank's deposit accounts.
 (a) single agent in each class; (b) random connectivity, with 2 banks, 3 firms, 4 workers.}
\label{graph.bmw}
\end{figure}

A graphical representation of monetary transactions with randomly sampled 
connections between agents is given in Fig. \ref{graph.bmw}. 

Because $\mathbf{A}_{nf,nh}, \mathbf{B}_{nf,nf}, \mathbf{C}_{nf,nh}, 
\mathbf{D}_{nb,nf}, \mathbf{E}_{nb,nh}$ are realizations of random variables,
$\xi$ belongs to an ensemble of random matrices that will be noted $\Xi(nb,nf,nh)$.

When $\xi$ is constant, $S$ defines a convex bounded polytope, that appears
in many disciplines, as will be discussed in section \ref{sub.related}.
The solutions to this class of under-specified problems 
will be examined in sections \ref{sub.csp} and \ref{sub.har}.
The reader interested in the study of other explicit specifications of flows may 
look at \cite{keen_solving_2010} and \cite{keen_monetary_2013}. Lavoie and Godley 
examine many increasingly detailed models (open economy,...), as well as dynamic specifications of
the flows\footnote{~see an implementation \url{https://github.com/kennt/monetary-economics}}.

To the best of our knowledge, the properties of random matrices that correspond to SFC 
models have not been established in the economics litterature. This representation 
calls for a closer examination following many works in the field of complex networks
studies.

\subsection{Related works in economics, econophysics, and network science}
\label{sub.related}
Double-entry book-keeping is used to establish National Accounts 
as a means to record estimated flows and stocks consistently,
for each country, at several levels of aggregation. National accountants 
build balance sheets, income and product accounts, in order to compare 
economies, and to analyze the behavior of economies in time, such as 
growth.
Such accounts are also used in macroeconomic modeling by policymakers 
to calibrate Computable General Equilibirum models. However, models of
this type (e.g. DSGE) do not explicitly enforce stock-flow consistency.  

Long before the break of the subprime crisis, many works have established 
ABM as an alternative, in a move to get rid of hypotheses
perceived as unjustified. Equilibrium, rationality, and the 
hypothesis of the representative agent have been criticized by many
economists \cite{kirman_complex_2011, fagiolo_macroeconomic_2012}.

Even though all ABM do not verify stock-flow consistency, it is an
hypothesis used by several authors \cite{dawid_agent-based_2008,
dawid_eurace_2011,seppecher_flexibility_2012,kinsella_income_2011,bezemer_causes_2011}. 

ABM are praised for their flexibility, their ability to
study large populations of heterogeneous and learning agents that interact
in possibly non-linear ways.
However, ABM need computer intensive simulations and face the problem of
being over-parametrized. Calibration is thus difficult and unstable, all
the more since empirical studies and reliable data are not abundant compared
to the dimension of the parameter space. Recent works tackle the issue of 
an efficient exploration of the parameter space \cite{salle_efficient_2014}.
The issue of comparing a model to experimental data is another problem posed to
practitioners, that has been addressed by different methods \cite{guerini_method_2016}.  
Under simplifying assumptions such as the aggregation of a subset of agents,
theoretical results concerning some macro-economic ABM were recently obtained
in a stock-flow-consistent framework, and phase diagrams established,
for specific dynamic rules \cite{gualdi_tipping_2015, gualdi_monetary_2015}. 

The importance of the organization of interactions, embodied by networks,
has been stressed in economics \cite{jackson_social_2013}. Theoretical works have 
studied their generic properties (supply chain \cite{weisbuch_production_2007},
interbank network \cite{delpini_evolution_2013}, trade credit 
\cite{battiston_credit_2007}). Empirical studies have laid emphasis
upon the topology of real economic networks, such as goods market,
national inter-firm trading \cite{watanabe_biased_2012}, world trade 
\cite{garlaschelli_fitness-dependent_2004}, global corporate control 
among transnational corporations \cite{vitali_network_2011},
firm ownership networks. They must sometime be reconstructed, starting
from limited information (of equity investments in the stock market 
\cite{garlaschelli_scale-free_2005}, the interbank market \cite{de_masi_fitness_2006}). 
At the level of individuals, detailed topological
information about banking, employment, or consumption seems to be lacking.

Such empirical and theoretical material may serve as an input to shape
the set of equations and inequalities discussed above.

\subsection{Constraint satisfaction problem}
\label{sub.csp}

As shown in section \ref{sub.bmw model}, the steady-state solution space associated with 
eq.(\ref{eq.matrix}) is a bounded convex polytope. Without loss of generality, $S$ 
can be supposed to have full row rank. The polytope is a $n-m$ dimensional
 object embedded in an $n$-dimensional space. 
 Properties of convex polytopes such as their different representations
are well studied  \cite{avis_lrs:_2000}. Computing their volume exactly can be
achieved by solving the vertex enumeration problem, which is $\#P$-hard.
Existing implementations, such as $lrs$ by Avis et al., allow to solve it
in reasonable time when $n-m$ is equal to 10 or below. 
Exact computation methods are employed in linear programming and operations
research to solve classical constraint satisfaction problems such as the map
coloring problem, and have real-life applications, for example in resource
allocation problems. More solutions can be obtained when relaxing hard
constraints.

Approximate methods to determine the solution space were proposed by
researchers studying metabolic networks and the metabolic steady-state flux
space. These methods allow to sample the solution space, to estimate its
volume, and to approximate probabilistic properties of the solutions such
as the marginal densities \cite{wiback_monte_2004,
 braunstein_estimating_2008,bianconi_flexibility_2006,
 bianconi_viable_2007,bianconi_flux_2008, massucci_novel_2013}. They
can be used to evaluate the sensitivity of the solution space to new
constraints.  A restriction of this problem known as Flux-Balance-Analysis
(FBA) consists in maximizing some objective constraint, which reduces
the solution space to a finite set of points \cite{orth_what_2010, bianconi_flux_2008}
or to an hyper-face. 
A parallel may be mentioned with the field of random Constraint
Satisfaction Problem (rCSP) that stands at the interface between theoretical
computer science and statistical physics, and studies sets of solutions
to a large number of random constraints, in a boolean space.
Many important results such as phase transitions were developed 
\cite{mezard_information_2009}. Although our main focus is the continuous domain,
we can take advantage of this theory.

\subsection{Monte-Carlo sampling of the steady-state solution space} 
\label{sub.har}
In section \ref{sub.csp} we mentioned the exact computation of the volume of
the steady-state solution space using vertex enumeration, when $n-m$ is small. 
The result obtained is numeric, and not an analytic expression depending 
on the parameters of the problem in eq. (\ref{eq.matrix}). 

Another approach proposed in \cite{wiback_monte_2004,almaas_global_2004},
suited for larger 
problems, is Monte-Carlo sampling in the solution space, using a
\emph{hit-and-run} algorithm. The latter needs an initial point inside the 
convex polytope, which can be found with a relaxation algorithm such
as MinOver \cite{krauth_learning_1987}. 
Then, sampling from a hypersphere, a direction is 
selected at random. The half-line defined by the starting point and
this direction intersects the boundary in a point. This intersection
and the starting point form a segment that can be uniformly sampled
to get the next point. The procedure defines a Markov Chain that 
converges to the uniform distribution over the polytope \cite{smith_efficient_1984,lovasz_hit-and-run_1998}, 
in nondeterministic polynomial time $\mathcal{O}^* (n^3 )$ after appropriate
preprocessing. 
The notation $\mathcal{O}^*()$ means that there are logarithmic factors that 
multiply $n^3$, and constants, but are neglected \cite{simonovits_how_2003}.

The \emph{hit-and-run} method provides an estimate of the marginal
probability density functions (pdf) denoted $P_i(x)$ for each unknown 
$i = 1,\dots,N$. Correlations between variables can also be estimated, as
well as other quantities that can be approximated with a finite sample of 
the solution space.
 They characterise the shape of the polytope and 
are given by: 
\begin{equation}
\label{eq.pdf}
P_i (x) = Vol(S_i (x))/Vol(S), ~S_i (x) = \{x \in  S ~s.t. ~x_i = x\}
\end{equation}
They can also be written as an integral over all stocks and flows. 
As remarked by \cite{wiback_monte_2004}, the \emph{hit-and-run} method 
will not permit us to estimate the absolute volume. Instead, approximations for relative volumes
can be obtained, such as $r=Vol(S_1)/Vol(S)$,  where $S$ is the same as in 
eq.(\ref{eq.matrix}) while $S_1$ has additional constraints, such as a lower bound for 
the variable $i$. 

To sum up, using efficient Monte-Carlo sampling methods such as 
\emph{hit-and-run}, we will be able to sample medium-sized problems (up 
to hundreds of variables), to approximate the pdfs, to compute relative  
volumes. However, we will get no analytic expression of theses quantities,
nor approximate entropy. Furthermore, the mixing of the Markov Chain should
be examined to ensure convergence. In section \ref{results} we use the implementation 
by Tervonen et al. \cite{tervonen_hit-and-run_2013}.

As evoked in section \ref{sub.csp}, researchers have also used the replica method 
\cite{bianconi_viable_2007},
and message passing algorithms \cite{braunstein_estimating_2008, 
font-clos_weighted_2012,bianconi_flux_2008} to deal with the problem of estimating
marginal densities. The computing time of message passing algorithms scales
as $\mathcal{O}(n)$ when the factor graph that represents the constraints 
contains no loop, but without this hypothesis convergence is not guaranteed. 

\section{Results}
\label{results}

\begin{figure}[htbp]
\centering
\includegraphics[width=14cm]{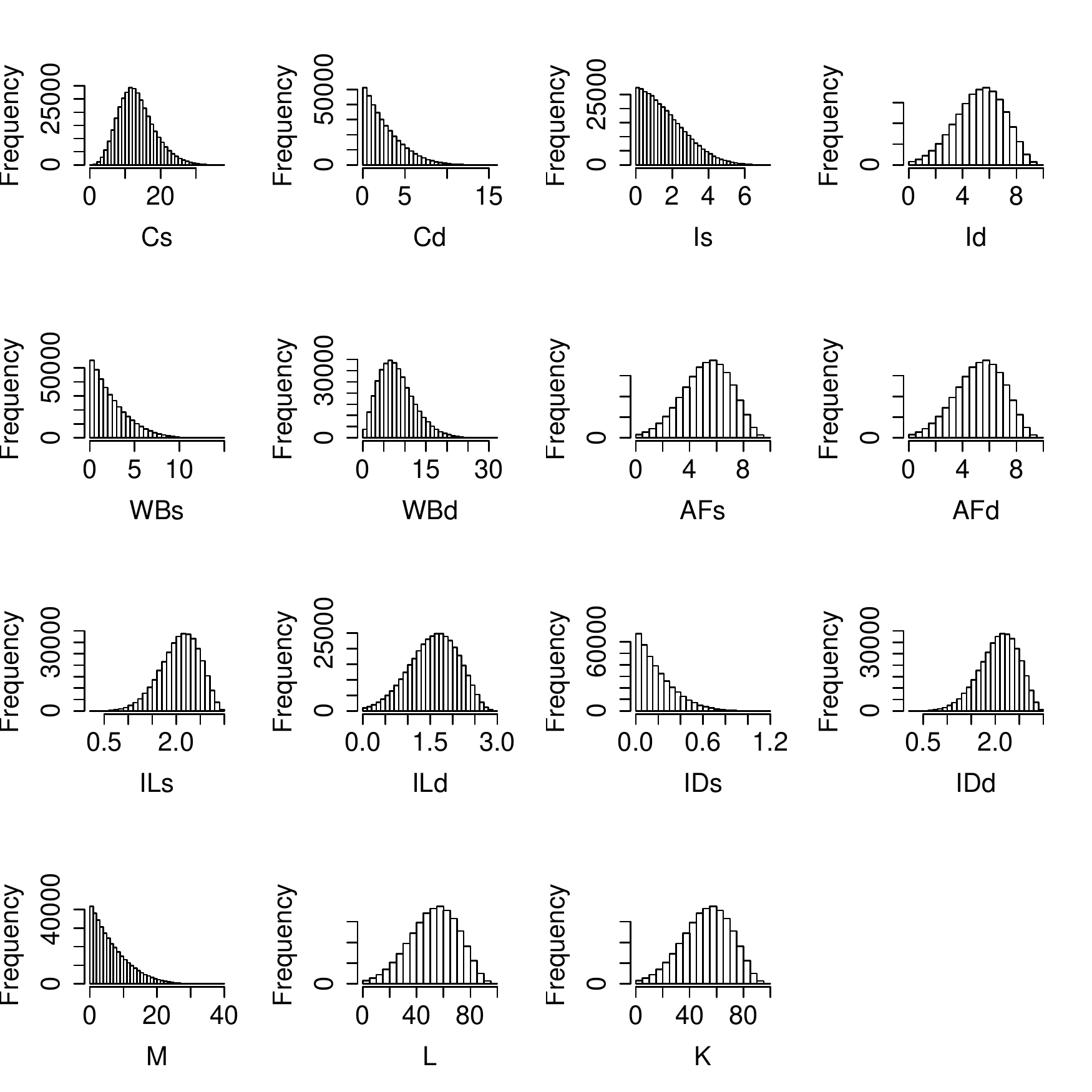}
\caption{Histograms corresponding to marginal pdf 
of individual steady-state monetary stocks and flows labelled according to
Tab.\ref{bmw short labels}, with $n_b=2, n_f=3, ~n_h=10$. The number of samples
solution of the constraint satisfaction problem is $N=3.10^5$. Subscripts
have been dropped for clarity. }
\label{histograms.simple}
\end{figure}

The marginal histograms of the variables of the constraint satisfaction
problem defined in section \ref{sub.manyagents}, where each sector is composed
of many agents, are represented in Fig. \ref{histograms.simple}. 
The solutions were sampled as explained in section \ref{sub.har},
with $\xi$ constant. The histogram are computed over the solutions, selecting
one agent in each sector.
The effect of averaging over $\Xi(nb,nf,nh)$ will be examined 
in section \ref{sub.randomize.matrix}.
The total quantity of deposits is constant, and the parameters have values summarized
in Tab. \ref{bmw params}.

We can first remark that the pdfs in Fig. \ref{histograms.simple} can be grouped
by shape: the pdf of $C_d$ and $WB_s$ that both appear in eq.(\ref{eq.cons}) 
have an exponential shape. The distribution of $ID_s$ has the same shape as that of $M$.
This is consistent with the linear relation in eq.(\ref{eq.interest.deposit}). 

\begin{figure}[htbp]
\centering
\subfigure[~]{
	\includegraphics[width=6cm, height=6cm]{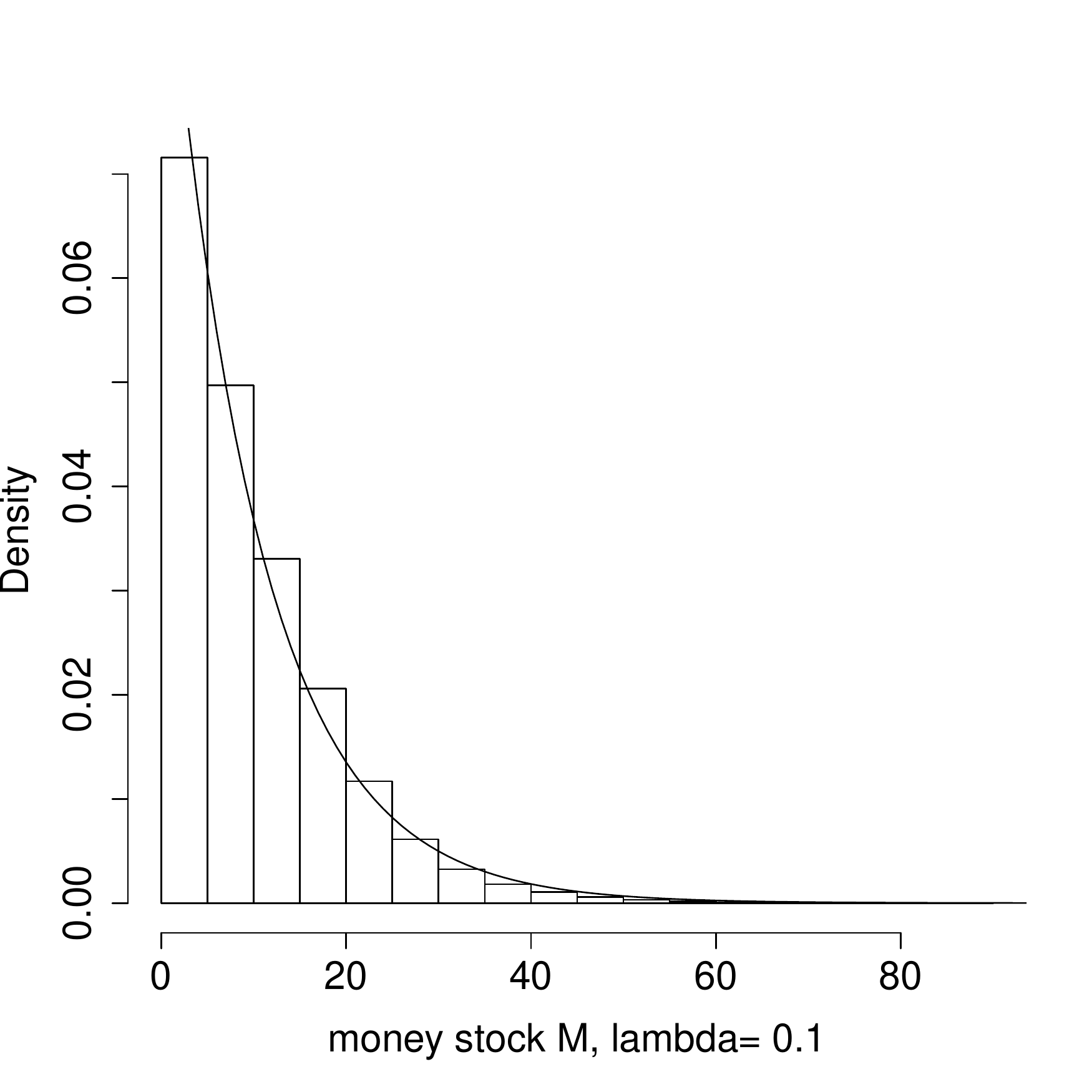}
	}
\subfigure[~]{
	\includegraphics[width=7cm]{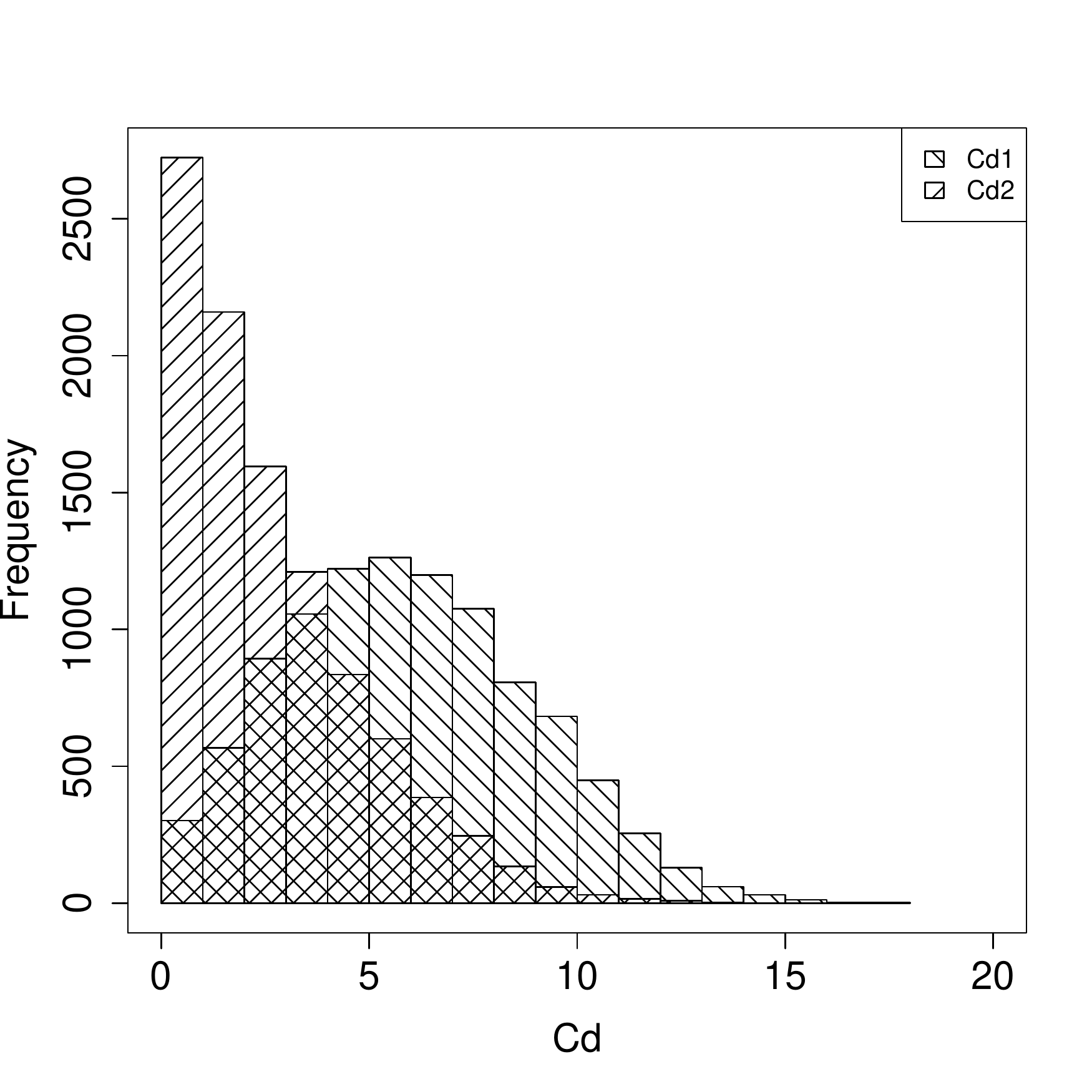}
	}
\subfigure[~]{
	\includegraphics[width=6cm,height=4cm]{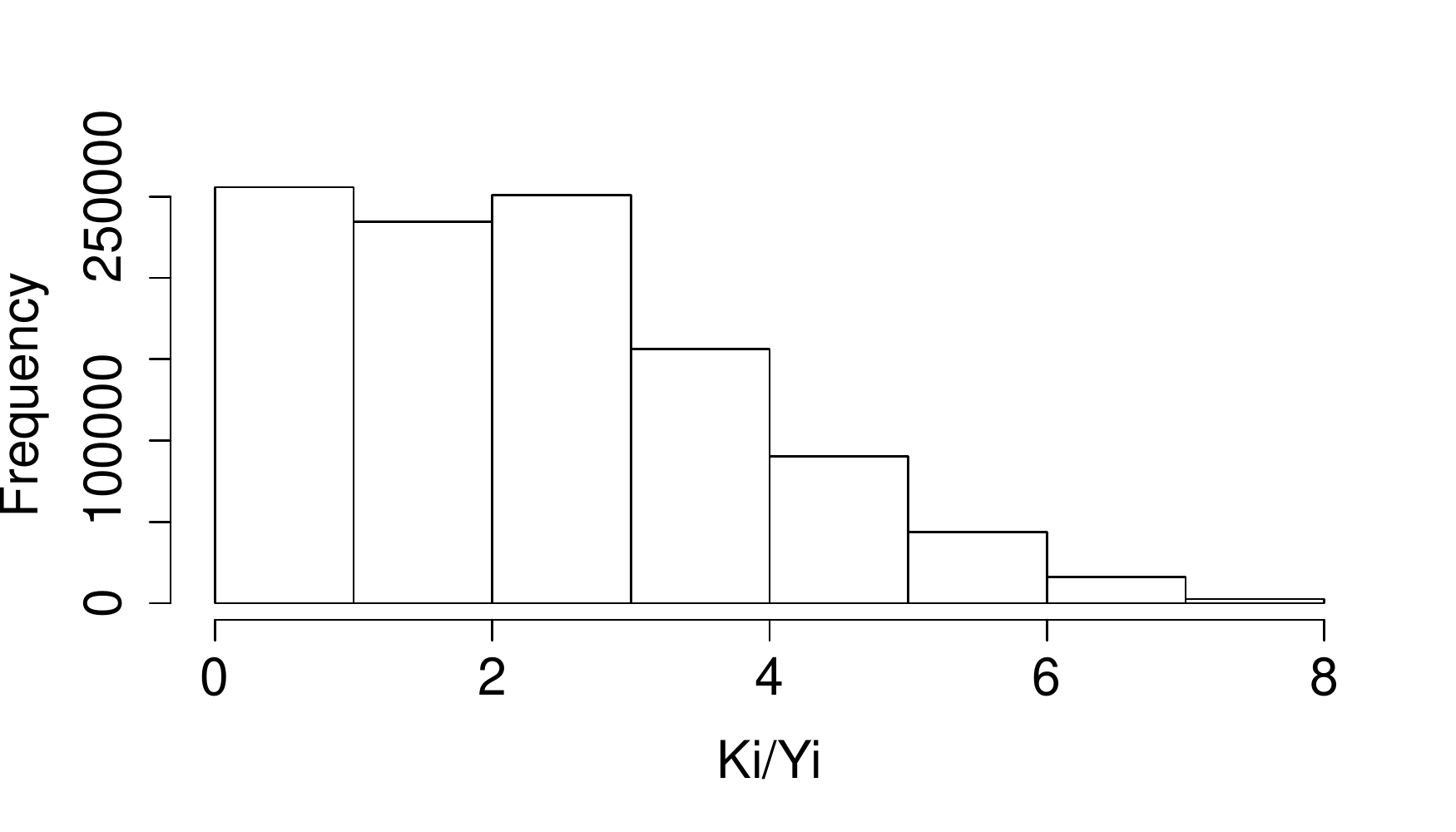}
	}	

\caption{ (a) Histogram of money stock $M$ and exponential fit; (b) histogram of $C_d$ for different agents;(b) capital output ratio.}
\label{fig.exp fit+ratio}
\end{figure}

The distribution of $M$ is well fitted by a continuous exponential distribution
as shown in Fig.\ref{fig.exp fit+ratio}(a). 
This can be compared with the empirical finding that, for many industrialized countries,
the lower part of the distribution of the wealth can be approximated by 
an exponential or gamma distribution \cite[2.3]{chakrabarti_econophysics_2013}.
The influence of re-sampling $\xi$ on this result will be examined
in section \ref{sub.randomize.matrix}.

The pdf of the stock of capital $K$ has the same shape as $I_d, AF_s, AF_d$ which are all
related by linear equations. The pdfs of $L$ and $IL_d$ have similar shapes, consistently
with eq.(\ref{eq.interest.loan}). This last shape will be discussed in
section \ref{sub.randomize.matrix}. 

A comparison between supply-side and demand-side can be made.
The mean flow of consumption supply is greater than the mean flow of consumption demand, 
in agreement with the fact that the number of firms is smaller than the number of
households.
Similarly, the mean wage supply $WB_s$, which goes to households, is lower than
the mean flow $WB_d$ paid by firms.  
The interest on deposit $ID_s$ paid to households has a lower mean value than the
$ID_d$ paid by banks, since the number of banks is smaller than the number of 
households.

Because the supply $C_s$ is for each firm the sum
of the demands of their clients, by a central limit argument
we can expect $C_s$ to converge to a normal law when the number of clients grows.

Because of topological effects, the distributions of flows across agents are 
heterogeneous, as examplified by the demand for consumption of two agents 
in Fig.\ref{fig.exp fit+ratio}(b).

The capital to output ratio $\gamma=K_i/Y_i=K_i/(C_{s,i}+I_{s,i})$ can be computed 
and its histograms over all firms and all solutions to the CSPs is shown
by Fig.\ref{fig.exp fit+ratio}(c). Econometric studies at the aggregate level report that
the value of this ratio has remained between 2 and 8 for many countries, on the long run
(see \cite{piketty_capital_2014} and the corresponding supplementary material\footnote{\url{http://piketty.pse.ens.fr/fr/capitalisback}}).

More generally, these remarks show that the model depicted in this article 
has interesting properties, can be easily modified, but needs more work
before being able to reproduce stylized facts. We discuss this topic in
section \ref{discussion}.

\begin{table}[htbp]
\centering
\begin{tabular}{cc}
\hline
Parameter & Value  \\
\hline
$\alpha_0$ & 0.0\\
$\alpha_1$ & 0.75\\
$\alpha_2$ & 0.1\\
$r$ & 0.03\\
$\delta$ & 0.1\\
$M_{tot}$ & 100 \\
\end{tabular}
\caption{Values of linear flows parameters, following \cite[chap. 7]{godley_monetary_2007}.}
\label{bmw params}
\end{table}

\subsection{The influence of knock-outs on systems with many agents} 
\label{sub.knockout}

The method of relative volumes devised in section \ref{sub.har} 
is used to study the impact of an external perturbation on the volume
of the solution space, for example when a flow is constrained.
This is termed \emph{knock-out} in metabolic network analysis, 
but in the present context can be interpreted as a limitation imposed on 
monetary variables, such as a credit shortage. Let us define:  

\begin{eqnarray}
\label{eq.constraints}
S_I &=& \{X \text{s.t} ~\xi X= 0, ~0\leq x_i \leq x_i^{\max}, ~\forall i \in I \} \\
S^\alpha_{I\setminus I_0} &=& \{X \text{s.t}  ~\xi X= 0, ~0\leq x_i \leq x_i^{\max}, ~\forall i \in I\setminus I_0, 
		 ~0\leq x_i \leq \alpha x_i^{\max}, ~\forall i \in I_0 \} 
\end{eqnarray}
where $X=[x_1,\ldots,x_{n}]$, $n$ is the number of columns of $\xi$,
$I$ is the set indexing the space of variables, and $I_0$ the set of variables that will 
be partially knocked-out, by an amount $\alpha$. 
Let us also define: 
\begin{eqnarray}
\label{eq.def}
r_V&=&Vol(S^\alpha_{I\setminus I_0})/Vol(S_I) 
\end{eqnarray}
The volume ratio $r_V$ measures the impact of selective variable knock-out 
on the solution space.

\begin{figure}[htbp]
\centering
\subfigure[~]{
	\includegraphics[width=8cm]{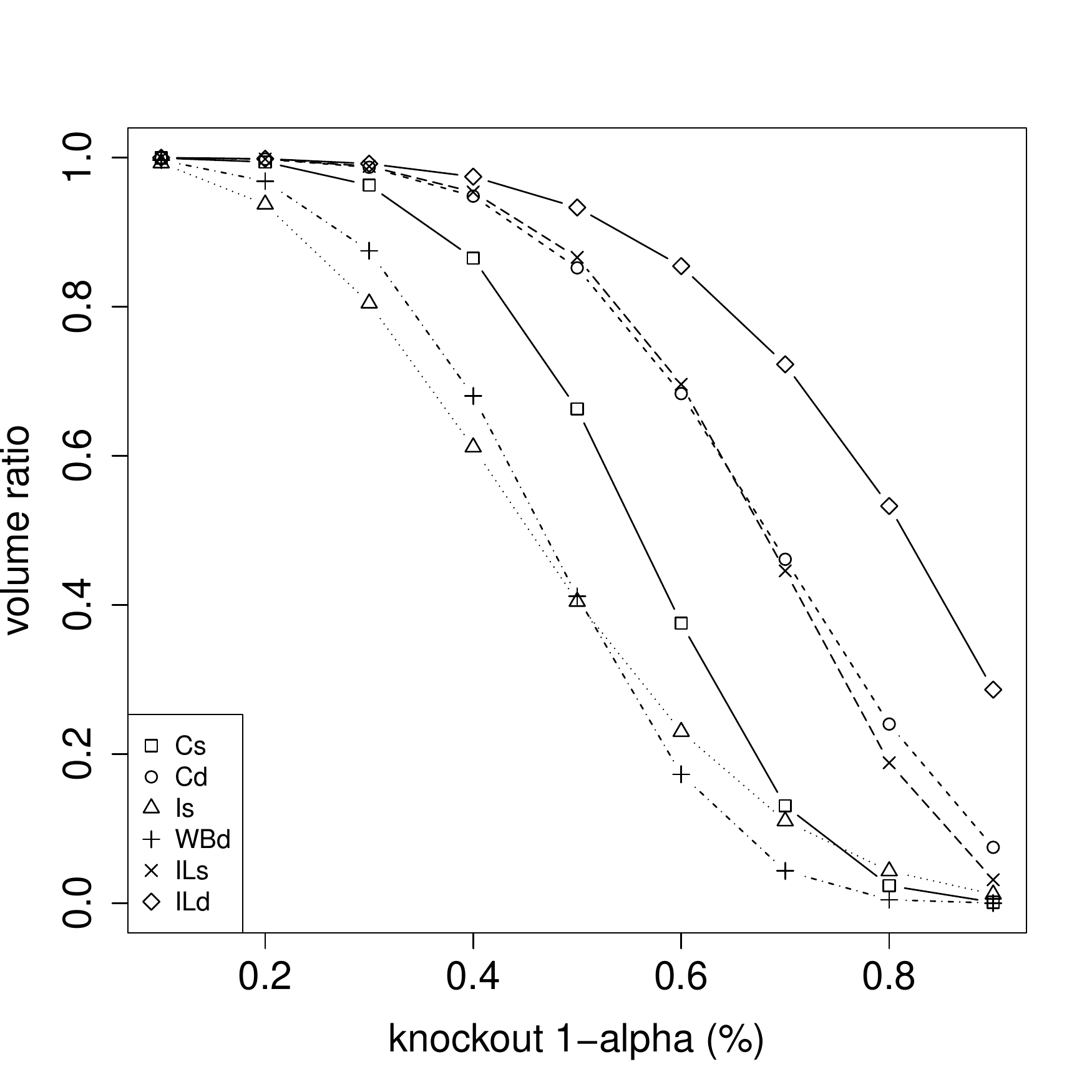}}
\subfigure[~]{
	\includegraphics[width=8cm]{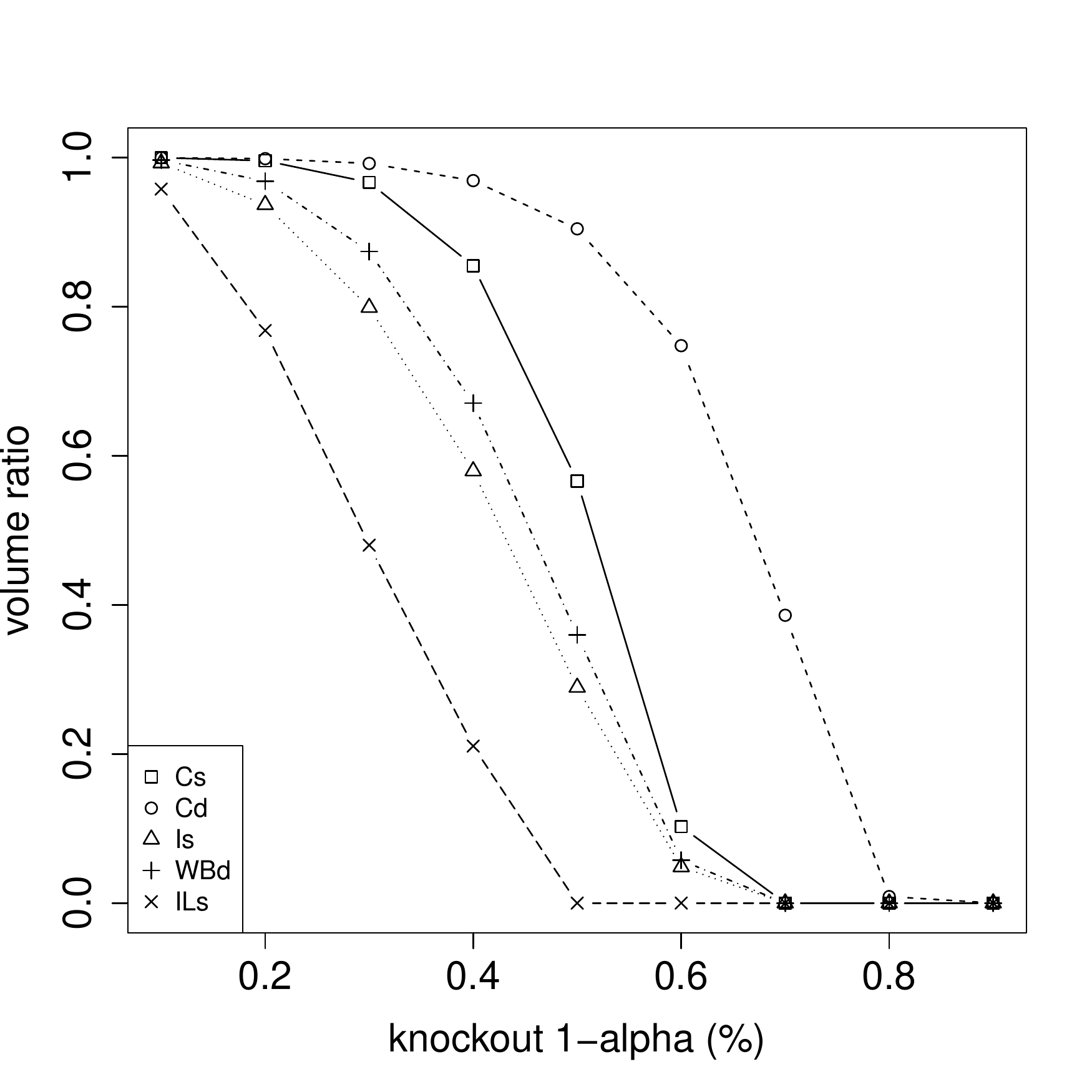}}
\caption{Impact of knock-out on volume ratios, in the  
case of a single draw of the random connectivity model depicted 
in sec.\ref{sub.bmw model}, with 2 banks, 3 firms and 10 workers.
Volume ratio appear as functions of $(1-\alpha)$. 
(left) single flow knock-out; (right) group flow knock-out.
Flow labels are explained in Tab. \ref{bmw short labels}. 
The number of sampled points is $n=3.10^5$. }  
\label{fig.knockout}
\end{figure}

We sample one random topology, as shown in Fig. \ref{graph.bmw}(b), 
with 2 banks, 3 firms and 10 households. Keeping this topology fixed for the 
rest of this section, we sample the solution space and summarize the 
results in Fig. \ref{fig.knockout}, where superposed curves have been removed. 
All ratios are functions of $1-\alpha$. 



It can be remarked in Fig.\ref{fig.knockout}(a) that single flow knock-outs
are able to reduce significantly the volume of the steady-state solution space. 
For example, a 33\% knock-out on investement made by one firm ($I_s$) 
out of the three defined in this simple model entails a cut by 20\% 
of the volume ratio $r_V$. 

The variables most influenced by group knock-outs in Fig.\ref{fig.knockout}(b) appear to be the 
interest on loans received by banks ($IL_s$), the investment of firms ($I_S$),
the payment of wages to workers by firms ($WB_d$). 
This ranking is partially conserved at the individual and group level, as shown by Fig. \ref{fig.knockout}(a,b).

The variable least influenced by group knock-outs is the demand for consumption 
goods ($C_d$). The variables with significant marginal probability on the right 
of $[0,x_i^{\max}]$ (such as $IL_s$ in Fig.\ref{histograms.simple}) undergo a 
large reduction because the tail is cut off. Conversely, variables 
with a small tail on the right of $[0,x_i^{\max}]$, such as $C_d$, show little reduction. 

Comparing the left and right panels of Fig. \ref{fig.knockout}, we remark that 
the influence of group knock-out on volume ratio is larger than single
knock-out. This remark is left for further theoretical analysis.

\subsection{Random matrix}
\label{sub.randomize.matrix}

So far, the matrix $\xi$ was considered constant. This hypothesis is interesting
when a particular economic network is examined.
However, when the emphasis is put more
on the properties of a random ensemble of networks than on a particular instance,
$\xi$ must be chosen randomly in $\Xi(nb,nf,nh)$.
Consequently, in this section, we consider empirical averages over both the solution
space and the set of random matrices $\Xi(nb,nf,nh)$. 

For computational reasons, the network size is set to $nb=2, nf=3, nh=6$.
Even in this particular case, the set of possible topologies is large, due to
its combinatorial nature, but can be sampled exhaustively in reasonable time.
Nevertheless, since for each randomly sampled matrix $\xi$ the corresponding solution space $S$
must also be sampled (which scales as $\mathcal{O}^* (n^3 )$, as discussed
in section \ref{sub.har}), the number of sampled matrices will be limited to one hundred. 

\begin{figure}[htbp]
\centering
\includegraphics[width=14cm]{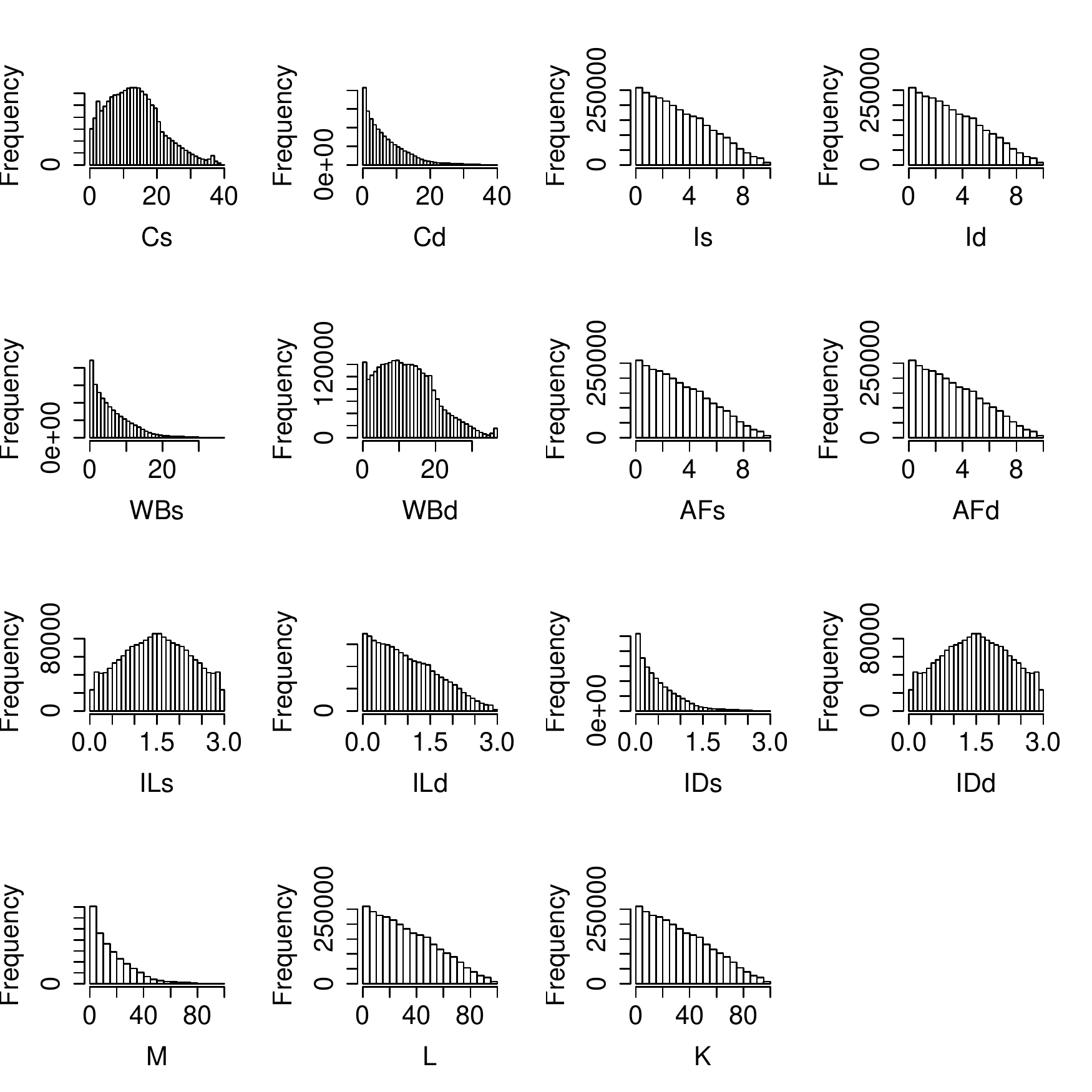}
\caption{Histograms corresponding to marginal pdf 
of steady-state monetary stocks and flows when $\xi$ is randomly sampled in
$\Xi(nb,nf,nh)$, with $nb=2, nf=3, ~nh=6$. $100$ random connectivity
matrices were generated. For each one, $10000$ hit-and-run solutions were sampled.}
\label{histograms.random}
\end{figure}

As illustrated by Fig. \ref{histograms.random}, the main observations made at the beginning of
section \ref{results} are recovered: the marginal distributions are grouped in the
same way, the difference between mean supply and demand is conserved, and the 
money deposit $M$ still has an continuous exponential shape.

We expect some effects related to topology to be averaged out,
such as the heterogeneity between agents, that was examplified in 
Fig.\ref{fig.exp fit+ratio}(b). However, to check this with the requested statistical
significance, we need to increase the number of sampled matrices, with an
acceptable computational load, as will we discussed in section \ref{discussion}.


\section{Discussion}
\label{discussion}



The first point we want to emphasize is that both the model and results 
in the sections above are preliminary. We do not claim that they can be of any
use at the moment regarding economic analysis. More work is needed, in collaboration
with economists, to improve their design, to evaluate the empirical stylized
facts they can reproduce, and eventually to validate the 
model \cite{guerini_method_2016} using empirical data.
In our view, the capacity to account for economic crises
can be seen through the rapid decay of the volume ratio in response to modest
fluctuations of the parameters, such as an increase in the knock-out factor. 
This can be seen independently of dynamical considerations.

In section \ref{sub.har} we presented an algorithm to compute
the volume of the solution space that belongs to the class of Monte-Carlo methods,
and illustrates the uniform sampling property of the \emph{hit-and-run} strategy. 
We didn't discuss convergence 
issues here, but they will become important as the size of the problem increases.
Furthermore, alternative estimation methods exist, for example
belief propagation that scales almost linearly in system size, and
can be applied when the variables are continuous \cite{font-clos_weighted_2012,fernandez-de-cossio-diaz_fast_2016}.

None of these algorithms provide litteral expressions that could be compared 
to results presented in section \ref{results},
such as the expression of absolute or relative volumes. 
For example, the magnitude of volume reduction provoked by knock-outs is
an important quantity from a system-scale point of view.
Interestingly, some theoretical results have been developped 
in the statistical physics literature evoked earlier, and should be compared to
our numerical experiments, in future works.

Concerning the complexity of the financial SFC model depicted in section \ref{sub.bmw model}, 
we started with a very simple setting, with a limited number of transaction types,
one agent per class, no state nor central bank. 
Flows were constrained in magnitude and depended on stocks. 
Then, the number of agents increased in section \ref{sub.manyagents} which gave us
a hint of the influence of topology. The accuracy of the model could be increased if 
random connectivity matrices were sampled in a random ensemble that corresponds to 
connectivity patterns observed empiricaly.
Seeking inspiration in macro-economic literature (e.g. Lavoie and Godley), we may
also add financial constraints on debt ratios at various levels.
Transactions not covered in this article can also be added, related to bonds, 
investments, in an open economy. A major improvement would be to 
couple the financial side to a production model of the economy, in order 
to compute prices, demand, unemployment and profit \cite{bardoscia_complexity_2015},
but many difficulties can be expected in that direction because of nonlinear relations that transform
the linear CSP into a nonlinear one. 

The issue of employment can't be addressed directly in this model because 
all households get a wage from some firm. Unemployed households should be
disconnected from firms, and included into another economic circuit, but this
is not a feature of the BMW model, nor of the extension proposed here. What
can still be studied thanks to the random nature of income in this framework,
is the proportion of households whose income is below a given threshold, which
could be related, for example, to aggregate demand.

In section \ref{results}, following \cite[chap. 7]{godley_monetary_2007}, 
with the hypothesis that flows are linear functions of stocks, we obtained 
interesting marginal quantities, such as the capital to output ratio. 
But this raises the issue of choosing the right parameter values. 
Various strategies can be implemented to address it, such as setting the parameters
according to empirical data taken from public statistics, or efficiently
exploring the space of parameters \cite{salle_efficient_2014}.
Another point of view is to consider the linear flows parameters as variables defined in specific 
intervals, and to include them in the sampling scheme used above. Although flows
are linear functions of stocks, this problem is also a nonlinear CSP, harder
to deal with than a linear one.

On the computational complexity side, a comparison should be made with other
classical approaches such as DSGE models and ABM with respect to the number
of agents, the sparseness of the network topology, the number of regions or 
countries, the richness of the financial mechanisms involved.
We can remark that in the case of \emph{hit-and-run} sampling, the main cost
is polynomial in the system size during sampling. Then in order to compute all the 
ratios discussed above, it is not necessary to resample: basic thresholding
is sufficient, and is linear in system size.

Furthermore, as reported in the random network community 
\cite{coolen_constrained_2009,squartini_unbiased_2015}, sampling  
network matrices $\xi$ with hard topological constraints raises the issue
of bias and efficiency, and will have to be controlled for in future works.

\section{Conclusion}
\label{conclusion}

We proposed an original strategy to compute macro-economic stocks and flows in a financial 
economy, inspired by stock-flow consistent models and methods developed in 
the field of metabolic networks. We show that this approach can be efficiently 
transposed thanks to approximate Monte-Carlo algorithms designed 
to solve Constraint Satisfaction Problems.
The steady-state in variable space is seen as a polytope included in a large 
dimensional space, which weighs equally all configurations of the financial
stocks and flows that are consistent with the constraints enforced by double-entry 
accounting. 

We proposed a random connectivity extension of the linear flow model
BMW by \cite{godley_monetary_2007}, 
and have obtained a numerical approximation of the probability density of 
stocks, flows, and the capital to output ratio. 
The money stock of households was found to be exponential in its lower part, which is
reminiscent of standard works in econophysics \cite{chakrabarti_econophysics_2013}. 

Flow knock-outs can be used to model economic phenomena such as credit shortages. 
We show that different types of constraints have distinct effects on the 
volume of the solution space, that can be interpreted as characterizing 
the flexibility of the financial flows. Rapid decay of the volume ratio
in response to modest fluctuations of the parameters can be interpreted
as crises, independently of dynamical specifications. 

Inside the class of SFC models, our approach fills a gap between SFC-ABM on one hand, 
and aggregate SFC models on the other hand, because the system size can scale
up to thousands of agents while preserving the possibility of a theoretical analysis,
and heterogeneity among agents. 
This comes at the cost of a simplification of the model, notably the hypothesis of
non-equilibrium stationary state, with linear flows. We discussed many potential 
improvements concerning the algorithms, the complexity of the model, and the relation to 
empirical data, and will deal with it in future works.




\appendix

\section{Acknowledgements}
The author wishes to thank two anonymous reviewers for their helpful comments.

Open-source software were used to perform this research: Python, R,
the R \emph{hit-and-run} package \cite{tervonen_hit-and-run_2013}, \emph{lrs},
graphviz, pygraphviz.

\bibliographystyle{elsarticle-num} 
\bibliography{HRF+eco}





\end{document}